\documentstyle[preprint,aps]{revtex}
\begin{document}
\draft
\date{\today}
\title{Comparison of Variational Approaches for the Exactly Solvable
1/r-Hubbard Chain}

\author{Florian Gebhard and Andreas Girndt}
\address{Dept.~of Physics and Materials Sciences Center, Philipps University,
35032~Marburg, Germany}

\maketitle%
\begin{abstract}%
We study Hartree-Fock, Gutzwiller, Baeriswyl, and combined
Gutzwiller-Baeriswyl wave functions for the exactly solvable
one-dimensional $1/r$-Hubbard model. We find that none of these
variational wave functions is able to correctly reproduce the
physics of the metal-to-insulator transition which occurs
in the model for half-filled bands when the interaction strength
equals the bandwidth.
The many-particle problem to calculate the variational
ground state energy for the Baeriswyl and combined Gutzwiller-Baeriswyl
wave function is exactly solved for the~$1/r$-Hubbard model.
The latter wave function becomes exact both for small and large
interaction strength, but it incorrectly predicts the metal-to-insulator
transition to happen at infinitely strong interactions.
We conclude that neither Hartree-Fock nor
Jastrow-type wave functions yield reliable
predictions on zero temperature phase transitions
in low-dimensional, i.e., charge-spin separated systems.

\end{abstract}
\pacs{PACS1993: 71.27.+a, 71.30.+h}

\narrowtext
\section{Introduction}
\label{sec1}
Variational wave functions are often used to study
ground state properties of quantum mechanical many-particle systems.
Examples of approximate ground state wave functions
for correlated Fermions are Hartree-Fock~\cite{Penn},
Gutzwiller~\cite{Gutzwiller}, local Ansatz~\cite{Fulde},
and, in general, Jastrow-Feenberg wave functions~\cite{Feenberg}.
One of the advantages of variational wave functions is the fact
that they give an exact upper bound for the true ground state energy.
Thus there is, at least in principle, a
criterion to assess variational wave functions:
the lower the variational ground state energy, i.e, the tighter
the variational bound, the better the
wave function (``energy criterion'').

In practise, however, two major problems arise:
(i)~the energy might not be a good criterion for the ``quality''
of a wave function, i.e., a wave function might give
a very good approximation for the ground state energy but it may
still miss the ground state physics of the system under consideration.
This issue can, however, only be addressed, if the exact solution
of the corresponding Hamiltonian is known;
(ii) the evaluation of correlated wave functions
poses yet another quantum many-particle
problem such that even seemingly simple wave functions
(and their straightforward improvements)
are not analytically tractable.

To deal with these two problems we focus on the exactly solvable
one-dimensional $1/r$-Hubbard model~\cite{prl}.
It describes spin-$1/2$ Fermions on a chain of $L$ sites, hopping with
long-range amplitude, $t_{l,m} = it (-1) ^{l-m} \left[d(l-m)\right]^{-1}
= t_{m,l}^{*}$.
Here, $d(l-m) = (L/\pi )\sin [\pi (l-m)/L]$ is the chord distance between
sites $l$ and $m$ on the chain closed into a ring
(the lattice spacing~$a$ is set to unity). The hopping becomes
$t_{l,m} \to it/(l-m)$ in the thermodynamical limit
$L \to \infty$ for fixed distance $(l-m)$ (``$1/r$-hopping'').
The electrons interact through a local Hubbard~\cite{Hubbard}
interaction, $U$,
and the total $1/r$-Hubbard Hamiltonian reads
\begin{equation}
\hat{H} = \hat{T} + U \hat{D} =
\sum_{l \neq m = 1,\sigma}^{L} t_{l,m} \hat{c}_{l,\sigma}^{+}
\hat{c}_{m,\sigma}^{\mbox{}}
+ U \sum_{l=1}^{L} \hat{n}_{l,\uparrow}\hat{n}_{l,\downarrow} \quad .
\label{hamilt}
\end{equation}
For even~$L$ we choose antiperiodic boundary conditions, so that
the resulting dispersion relation is linear in wave vector,
namely, $\epsilon (k) = t k$ with $k=\Delta (m+1/2)$ ($\Delta =2\pi /L$,
$m=-L/2,\ldots,L/2 -1)$. For $U=0$ the Fermi sea is the
ground state with all $k$-states from $k=-\pi$
to $k_F^e=\pi (n-1)$ filled where $n = (N_{\uparrow} +N_{\downarrow})/L$
is the total particle density.
For half-filling ($n=1$) and $U\to \infty$, the Hamiltonian~(\ref{hamilt})
reduces to the spin-1/2 $(1/r)^2$-Heisenberg or
Haldane-Shastry model~\cite{HSS}.
In the rest of the paper we will restrict ourselves to~$n \leq 1$.
The more than half-filled case can be obtained by particle-hole
symmetry~\cite{prb}.

The exact solution of the~$1/r$-Hubbard Hamiltonian
was conjectured in Ref.~\cite{prl}, and its physical
properties were discussed in detail elsewhere~\cite{prb}. The model displays
a Mott-Hubbard metal-to-insulator transition at half-filling
when the interaction strength,~$U$, equals the
bandwidth,~$W=2\pi t$. This is in contrast to the one-dimensional
Hubbard model which was exactly solved by Lieb and Wu~\cite{LiebWu}.
There, the tight-binding cosine dispersion has a perfect nesting
wave vector $q=\pi$ such that
$\epsilon_{\rm tb} (k) = -2t\cos k = -\epsilon_{\rm tb} (k+q)$.
Consequently, the Hubbard model describes an insulating state
for all $U>0$ at half-filling. This metal-to-insulator transition
at $U=0^+$ for $n=1$ is trivially reproduced in
an antiferromagnetic Hartree-Fock theory~\cite{Penn}.
We therefore use the
$1/r$-Hubbard model to assess the quality of the
Hartree-Fock approach for models {\em without} the perfect nesting
property.

Another advantage of the $1/r$-Hubbard model comes from the fact that
we are able to calculate variational ground state energies
for correlated wave functions without further approximations.
Thus far, this was only achieved for the Gutzwiller wave function
in one dimension~\cite{MV,DVFG,review}, or, for general Gutzwiller-correlated
wave functions, in the limit of large dimensions~\cite{MVlarged,FG1,Rainer} or
large orbital degeneracies~\cite{FG2}.
For the $1/r$-Hubbard model we are able to derive the
ground state energies for the Baeriswyl~\cite{Baeriswyl}
and a combined Gutzwiller-Baeriswyl wave function.

The plan of the paper is as follows: in section~\ref{sec2}
we diagonalize the antiferromagnetic Hartree-Fock Hamiltonian to obtain
the Hartree-Fock ground state wave function and energy.
In section~\ref{sec3} we introduce and
evaluate the Gutzwiller-, Baeriswyl-, and Gutzwiller-Baeriswyl
wave functions. We draw our conclusions
on the quality of these variational approaches in section~\ref{sec4}.

\section{Hartree Fock Approximation}
\label{sec2}
As usual~\cite{Penn} we factorize the interaction part of the
Hamiltonian~(\ref{hamilt})
as
\begin{mathletters}%
\begin{equation}%
 \hat{n}_{l,\uparrow}\hat{n}_{l,\downarrow} \rightarrow
 \langle \hat{n}_{l,\uparrow} \rangle \hat{n}_{l,\downarrow}
 + \hat{n}_{l,\uparrow} \langle \hat{n}_{l,\downarrow} \rangle
 - \langle \hat{n}_{l,\uparrow} \rangle \langle \hat{n}_{l,\downarrow} \rangle
\label{HFfactorization}
\end{equation}
and allow for a sublattice magnetization
$m= |\langle \hat{n}_{l,\uparrow} \rangle - \langle \hat{n}_{l,\downarrow}
\rangle|$, alternating in sign on the odd/even lattice points
of the $A$-$B$ lattice. The particle
density~$n=\langle \hat{n}_{l,\uparrow} \rangle + \langle
\hat{n}_{l,\downarrow}
\rangle $ remains uniform. Thus we may write
\begin{equation}
\langle \hat{n}_{l,\sigma} \rangle = \frac{n}{2} + (-1)^l \sigma \frac{m}{2}
\quad .
\end{equation}%
\end{mathletters}%
We introduce the new quasiparticle operators~$\hat{a}_{k,\sigma}$,
$\hat{b}_{k,\sigma}$ within the magnetic Brillouin zone $-\pi < k < 0$
by
\begin{mathletters}%
\begin{eqnarray}%
\hat{a}_{k,\sigma}& =& \cos \, \phi_{k,\sigma} \, \hat{c}_{k,\sigma} +
\sin \phi_{k,\sigma} \, \hat{c}_{k+\pi,\sigma}\nonumber \\
\hat{b}_{k,\sigma}& =& -\sin \phi_{k,\sigma} \,\hat{c}_{k,\sigma} +
\cos \phi_{k,\sigma} \, \hat{c}_{k+\pi,\sigma}
\end{eqnarray}
and the inverse transformation reads
\begin{eqnarray}%
\hat{c}_{k,\sigma}& =& \cos \phi_{k,\sigma} \, \hat{a}_{k,\sigma}
- \sin \phi_{k,\sigma} \, \hat{b}_{k,\sigma}\nonumber \\
\hat{c}_{k+\pi,\sigma}& =& \sin \phi_{k,\sigma} \, \hat{a}_{k,\sigma} +
\cos \phi_{k,\sigma} \, \hat{b}_{k,\sigma} \quad .
\end{eqnarray}%
\end{mathletters}%
The band width of our model is~$W =2\pi t$, and the transformation
angle~$\phi_{k,\sigma}$ fulfills
\begin{mathletters}%
\begin{eqnarray}
\tan 2 \phi_{k,\sigma} & = & \sigma \frac{2Um}{W} \\[6pt]
\sin^2 \phi_{k,\sigma} &= & \frac{1}{2}
\left( 1 - \sigma \frac{W}{\sqrt{W^2 +(2 U m)^2}} \right)
\quad .
\end{eqnarray}%
\end{mathletters}%
The operators are constructed such that $\hat{a}_{k,\sigma}^+$
($\hat{b}_{k,\sigma}^+$) creates particles mainly on
the~$A$-lattice ($B$-lattice) sites. The Hartree-Fock Hamiltonian then reads
\begin{mathletters}%
\begin{equation}%
\hat{H}_{\rm HF} = \sum_{-\pi < k < 0}
E_{k}^- \left( \hat{a}_{k,\uparrow}^+ \hat{a}_{k,\uparrow}^{\phantom{+}}
+ \hat{b}_{k,\downarrow}^+ \hat{b}_{k,\downarrow}^{\phantom{+}} \right)
+
E_{k}^+ \left( \hat{a}_{k,\downarrow}^+ \hat{a}_{k,\downarrow}^{\phantom{+}}
+ \hat{b}_{k,\uparrow}^+ \hat{b}_{k,\uparrow}^{\phantom{+}} \right)
\label{hamiltHF}
\end{equation}
with
\begin{equation}
E_k^{\pm} = tk +\frac{1}{4} \left( W + U + Um^2\right)
\pm \frac{1}{4} \sqrt{W^2 + 4 U^2 m^2 } \quad .
\end{equation}%
\end{mathletters}%
The Hartree-Fock charge gap is given by~$\Delta\mu_c^{\rm HF} = E_{k=-\pi}^+
-E_{k=0}^- = \left[ \sqrt{W^2 + 4 U^2 m^2 } -W\right]/2$.
The corresponding ground state wave function reads
\begin{equation}
| \psi_{0}^{\rm HF} \rangle = \prod_{-\pi < k < k_{F}^{\rm HF}}
\hat{a}_{k,\uparrow}^+ \hat{b}_{k,\downarrow}^+ |{\rm vacuum}\rangle
\quad .
\end{equation}
Finally, we obtain the particle densities for all temperatures~$T=1/\beta$
as
\begin{mathletters}%
\begin{eqnarray}%
\langle \hat{n}_{k,\uparrow}^a \rangle &=&
\frac{1}{\exp \left( \beta (E_k^- -\mu_{\rm HF}) \right) +1}
= f_{\rm FD} (E_k^-) = \langle \hat{n}_{k,\downarrow}^b\rangle \\[6pt]
\langle \hat{n}_{k,\uparrow}^b \rangle &=&
\frac{1}{\exp \left( \beta (E_k^+ -\mu_{\rm HF}) \right) +1}
= f_{\rm FD} (E_k^+) =\langle \hat{n}_{k,\downarrow}^a \rangle
\end{eqnarray}%
\end{mathletters}%
where the Hartree-Fock chemical potential~$\mu_{\rm HF}$ is obtained from the
particle number
\begin{mathletters}%
\begin{equation}
n = \frac{1}{L} \sum_{-\pi < k < \pi,\sigma} \langle \hat{n}_{k,\sigma}\rangle
= \frac{2}{2\pi} \int_{-\pi}^0 dk \left[ f_{\rm FD} (E_k^-) + f_{\rm FD}(E_k^+)
\right] \quad .
\label{muHF}
\end{equation}
The self-consistency condition reads
\begin{eqnarray}
m &=& |\langle \hat{n}_{l,\uparrow} \rangle - \langle \hat{n}_{l,\downarrow}
\rangle| \nonumber\\[3pt]
&=& \frac{4 U m}{\sqrt{W^2 +4 U^2 m^2}}
\left| \frac{1}{2\pi} \int_{-\pi}^0 d k  \left[ f_{\rm FD} (E_k^-) -
f_{\rm FD}(E_k^+) \right]   \right| \quad .
\label{selfconsistency}
\end{eqnarray}%
\end{mathletters}%
This equation gives the magnetization as a function of temperature and
interaction strength.

We note that the Hartree-Fock Hamiltonian~(\ref{hamiltHF})
will give a {\em finite} magnetization at {\em finite} temperatures,
if $m>0$ at $T=0$. However, a phase transition from a paramagnetic to an
antiferromagnetic phase at some finite Ne\'{e}l temperature does not
happen in a one-dimensional
system~\cite{Uhrig}. This is one of the well-known
shortcomings of the Hartree-Fock Mean-Field approach to low-dimensional
systems.
Furthermore, it gives a $T=0$ transition to a magnetic phase
at a critical value~$U=U_c^{\rm HF}(n)$ for {\em all}
band-fillings because a finite magnetization is the only way to obtain
a lower energy than the paramagnetic Hartree-Fock phase for large~$U/W$.
We thus restrict ourselves to the half-filled case ($n=1$)
where the Hartree-Fock approach gives the best results.
For $n<1$ the antiferromagnetic Hartree-Fock theory certainly fails
to give the correct physical picture.

In the case of half-filling the Hartree-Fock chemical potential
follows from eq.~(\ref{muHF}) to be temperature independent.
It is given by $\mu_{\rm HF} = U\left(1+m^2\right)/4$
which lies in the middle of the gap between the upper and lower
Hartree-Fock band.
The self-consistency equation~(\ref{selfconsistency}) for the magnetization
does not considerably simplify but stays an implicit equation which has
to be solved numerically for $T>0$.
Here, we are only interested in ground state
properties. For $\beta \to \infty$ the lower Hartree-Fock band is completely
filled ($k_F^{\rm HF}=0$),
and it follows from~(\ref{selfconsistency})
that either~$m=0$ (paramagnetic Hartree-Fock solution) or
\begin{equation}
2 U  = \sqrt{W^2 +4 U^2 m^2} \quad .
\end{equation}
Hence, Hartree-Fock predicts a transition from a paramagnetic metal
to an antiferromagnetic insulator at $U_c^{\rm HF}(n=1) = W/2$,
above which we have~$m= \sqrt{1 - \left( U_c^{\rm HF} / U\right)^2}$
with the typical mean-field exponent of one-half near the transition.
The charge gap is
$\Delta\mu_c^{\rm HF}
= \left( \sqrt{W^2 +4 U^2 m^2} -W \right)/2 = U-U_c^{\rm HF}$.
The Hartree-Fock gap grows {\em linearly} as a function of~$U-U_c^{\rm HF}$
in agreement with
the exact solution~\cite{prl,prb} where, however, the true transition is
at {\em twice} the Hartree-Fock value, $U_c=2U_c^{\rm HF}$.
It is amusing to note that the density of states for spinon excitations
develops a van-Hove singularity (formation of the upper
and lower Hubbard band) precisely at~$U=W/2$~\cite{prb}.
It is again seen that antiferromagnetic Hartree-Fock is sensitive to
spin-correlations only.

The Hartree-Fock variational ground state energy density at half-filling
is given by $e_0^{\rm HF} = (1/L) \sum_{-\pi < k < 0} E_{k}^-$
which gives
\begin{equation}
e_0^{\rm HF} = \left\{
\begin{array}{ccll}
{\displaystyle \frac{U-W}{4}} & {\rm for} & U\leq W/2 & (m=0) \\[9pt]
{\displaystyle -\frac{W^2}{16 U}}
& {\rm for} & U \geq W/2 & (m=\sqrt{1-\left(W/2U\right)^2})
\end{array}
\right.
\end{equation}
which is compared in fig.~\ref{fig1} to the exact ground state energy
\begin{equation}
e_0 = \left\{
\begin{array}{ccl}
{\displaystyle \frac{U-W}{4}- \frac{U^2}{12 W}} & {\rm for} & U\leq W \\[9pt]
{\displaystyle -\frac{W^2}{12 U}} & {\rm for} & U \geq W
\end{array}
\right.
\quad .
\label{e0exact}
\end{equation}
It is seen that Hartree-Fock, as an effective single-particle
theory, fails to give any contribution to the correlation energy
of order~$U^2/W$. There are no non-analytic contributions of the
order~$\exp (-W/U)$ for small~$U/W$ as expected for a Hamiltonian without
the perfect nesting property.

At large~$U/W$ it gives the correct
analytical behavior with a smaller prefactor.
However, it also predicts a finite sublattice magnetization
which is, of course, zero in the exact solution.
Furthermore, the Hartree-Fock gap is temperature dependent and
one gets a transition from an antiferromagnetic insulator to a
paramagnetic metal as function of temperature, if $U>U_c^{\rm HF}$.
The gap in the exact solution is temperature {\em independent},
and no transitions occur at finite~$T$. Finally, the zero temperature
transition itself in Hartree-Fock is driven by {\em spin-correlations}.
This can be seen from the fact that one has a finite sublattice magnetization
after the transition, and that the transition happens at all
fillings. This is in contrast to the exact solution
where the transition happens only at half-filling, and no magnetic
ordering occurs. Hence, the transition in the $1/r$-Hubbard model
is driven by {\em charge-correlations}, just opposite to the physics
of the Hartree-Fock approximation.

\section{Correlated Wave Functions}
\label{sec3}
We now focus on the correlated wave functions whose energy we want
to evaluate.
Our starting point is the Gutzwiller wave function~\cite{Gutzwiller}
\begin{equation}
| \psi_{\rm G} (g) \rangle = g^{\hat{D}} |\hbox{\rm Fermi-sea}\rangle
= \prod_{l=1}^{L} \left[ 1-(1-g) \hat{D}_{l} \right]
|\hbox{\rm Fermi-sea}\rangle
\label{GWF}
\end{equation}
which starts from the Fermi-sea, the exact ground state for~$U=0$ ($g=1$).
For $U>0$ double occupancies are less favorite because
of the on-site repulsion in the Hamiltonian, eq.~(\ref{hamilt}).
To include this effect, double occupancies are suppressed
by the Gutzwiller
correlator~$\prod_{l=1}^{L} \left[ 1-(1-g) \hat{D}_{l} \right]$ which
globally reduces configurations with
double occupancies in the Fermi-sea, if $g<1$.
Note that this correlator is regular
for all~$0<g\leq 1$, i.e., $| \psi_{\rm G} (g)\rangle$
describes a (correlated) metal as long as we do not project out
all double occupancies. For $g=0$ we alter the nature of the
Fermi-sea because we not only suppress configurations with
double occupancies but truly eliminate them with the help of
the Gutzwiller
projector~$\hat{P}_{D=0}=~\prod_{l=1}^{L} \left[ 1- \hat{D}_{l} \right]$.

A metal-to-insulator transition within this
wave function can only be expected at half-filling, if
the variational procedure gives~$g=0$ such that
every site is singly occupied (for less than half-filling
the remaining holes are still mobile and keep the state metallic).
Brinkman and Rice observed that such a transition
(``Brinkman-Rice transition''~\cite{BR,Vollhardtreview})
at a finite~$U=U_c^{\rm BR}= 8 |e_0^{\rm FS}|$
was contained in the Gutzwiller Approximation~\cite{Gutzwiller}
to the Gutzwiller wave function,
where $e_0^{\rm FS}$ is the kinetic energy of the Fermi-sea in the
non-interacting system.
It was later understood that
this approximation becomes exact in the limit of
infinite dimensions~\cite{MV,MVlarged}.
If we plainly {\em apply} the Gutzwiller Approximation to our one-dimensional
system we obtain
$e_0^{\rm FS}=-W/4$, $U_c^{\rm BR} = 2W$, and the ground state energy
density at half-filling in the Gutzwiller Approximation is given
by~$e_0^{\rm GA}= e_0^{\rm FS} (1-U/U_c^{\rm BR})^2$ for~$U\leq U_c^{\rm BR}$,
$e_0^{\rm GA}=0$ for~$U\geq U_c^{\rm BR}$.
This energy density
is plotted in figure~\ref{fig1} in comparison to the Hartree-Fock
and the exact result. Although
the Gutzwiller Approximation is an {\em uncontrolled}
approximation in any finite dimension, we obtain the
interesting result that it predicts a metal-to-insulator transition
at half-filling for a {\em finite} value~$U=U_c^{\rm BR}=2W$.
This transition is driven by charge-correlations only which is obvious
from the fact that the ground state energy density in this approximation
is zero beyond~$U_c^{\rm BR}$. It is thus seen that this approximation
very accurately describes the physical mechanism of the metal-to-insulator
transition in our one-dimensional model, despite the fact that
it is not variationally controlled, and the variational estimate for
the ground state energy density is poor for large~$U/W$.
In any finite dimension, the Brinkman-Rice transition will be shifted
to~$U_c^{\rm G}=\infty$ where double occupancies are strictly
forbidden by the Hubbard interaction.
This has indeed been proven for the Gutzwiller wave function~\cite{PvDetal}.

It was only recently that one was able to go beyond the Gutzwiller
Approximation to exactly evaluate the Gutzwiller wave function
in one dimension~\cite{MV,DVFG}. It was shown
that the Gutzwiller wave function does not give a good ground state energy
for the one-dimensional
Hubbard model for large~$U/W$ due to poor correlations
between double occupancies and holes~\cite{review}. On the other hand, the
correlations between spins are excellent such that the Gutzwiller
projected Fermi-sea is a very good trial state for the
one-dimensional antiferromagnetic Heisenberg model~\cite{KHF,DVFG,review}.
We will see below that these considerations remain valid for
the $1/r$-Hubbard model, eq.~(\ref{hamilt}).

Baeriswyl~\cite{Baeriswyl} constructed a variational
wave function for the
one-dimensional Hubbard model to overcome the difficulties
of the Gutzwiller wave function for large~$U/W$. He
started from the Gutzwiller-projected Fermi-sea to incorporate
its excellent spin correlations, and proposed
\begin{equation}
| \psi_{\rm B} (b) \rangle = b^{\hat{T}/W}
\hat{P}_{D=0} |\hbox{\rm Fermi-sea}\rangle
\label{BWF}
\end{equation}
as variational ground state ($0 \leq b \leq 1$).
For large~$U/W$ the variational parameter~$b$
is close to unity, and double occupancies and holes
are linked by the application of the kinetic energy operator to
the Gutzwiller-projected Fermi-sea. Hence,
the missing charge-correlations in the Gutzwiller wave function
are now properly taken into account~\cite{Baeriswyl}.

At half-filling, the Baeriswyl wave function describes an
insulator for all~$0<b\leq 1$ because the Gutzwiller-projected
half-filled Fermi-sea is insulating, and the
Baeriswyl-correlator~$b^{\hat{T}/W}$
is regular as long as $b>0$. A metal-to-insulator transition will
only happen, if $b=0$ is the result of the minimization procedure.
Since $b=0$ corresponds to the free Fermi-sea we can expect that
the transition will happen at $U=0^+$ for the Baeriswyl wave function.
Unfortunately, not much is known about the Baeriswyl wave function
because even the calculation of the variational
ground state energy is too difficult for general Hamiltonians.
Below, we show how this can be accomplished for the~$1/r$-Hubbard model.
Our analysis will confirm our general considerations on the
metal-to-insulator transition in this wave function.

Obviously,
it is necessary to go beyond Gutzwiller and Baeriswyl wave functions
because neither of them is able to describe the metal-to-insulator
transition of the $1/r$-Hubbard model at $U=W$. Their natural
generalization is the ``Gutzwiller-Baeriswyl'' wave function defined as
\begin{equation}%
| \psi_{\rm GB} (b,g) \rangle =
b^{\hat{T}/W} g^{\hat{D}} |\hbox{\rm Fermi-sea}\rangle
\quad .
\label{VWF}
\end{equation}
For $b=1$ this wave function reduces to the Gutzwiller wave
function~(\ref{GWF}),
and to the Baeriswyl wave function~(\ref{BWF}) for $g=0$.
We now have a two-parameter wave function which allows for
the competition of the ``kinetic'' Baeriswyl correlator
and the ``potential'' Gutzwiller correlator.
We might hope that this competition will result in a
variational prediction of a metal-to-insulator at some {\em finite} critical
value~$U_c^{\rm GB}$. We will see below that this is not the case.
Instead, we again find~$U_c^{\rm GB} =\infty$ for the $1/r$-Hubbard model.

We denote $\langle \hat{O} \rangle = \langle \psi_{\rm GB} | \hat{O} |
\psi_{\rm GB} \rangle / \langle \psi_{\rm GB} | \psi_{\rm GB} \rangle$.
Details of the calculation are presented in appendix~\ref{appa}.
The final result for the expectation value of the kinetic energy is
\begin{mathletters}
\label{resultVWF}
\begin{equation}
\langle \hat{T}(b,g) \rangle/L = -\frac{Wn(1-n)}{4}+\frac{W}{8}
\int_{-1}^{2n-1} dx \,
\frac{\beta(x)^2 - \alpha(x)^2}{\beta(x)^2 +\alpha(x)^2}
\end{equation}
and the mean double occupancy is given by
\begin{equation}
\langle \hat{D}(b,g) \rangle/L = \overline{d}(b,g)=
\frac{1}{8} \int_{-1}^{2n-1} dx \,
\frac{\left( \alpha(x)\sqrt{1+x} + \beta(x)\sqrt{1-x}\right)^2}%
{\beta(x)^2 +\alpha(x)^2}
\end{equation}%
\end{mathletters}%
where
\begin{mathletters}
\label{alphabeta}
\begin{eqnarray}
\alpha(x) & \equiv & \alpha (x;b,g) = \sqrt{\frac{1}{b}}
\left[ 1 +\frac{g-1}{2} (1+x) \right] \\[3pt]
\beta(x) &\equiv& \beta (x;b,g) = \frac{\sqrt{b}(g-1)}{2} \sqrt{1-x^2}
\quad .
\end{eqnarray}%
\end{mathletters}%
Before we proceed with the general case,
we first discuss the special cases $b=1$ (Gutzwiller wave function)
and $g=0$ (Baeriswyl wave function).

\subsection{Gutzwiller wave function}
\label{Gutz}
We set $b=1$ in eqs.~(\ref{resultVWF}), (\ref{alphabeta}), and
obtain $\alpha (x;g) = \left[ 1-x + g(1+x) \right]/2$, and
$\beta (x;g) = (g-1)\sqrt{1-x^2}/2$. After some straightforward manipulations
one arrives at
\begin{mathletters}
\begin{eqnarray}
\overline{d} (g) &=& \frac{g^2}{2} \int_{0}^{n} dy \,
\frac{y}{1 + (g^2-1)y}
\\[6pt]
\langle \hat{T}(g) \rangle/L &=& -\frac{Wn(2-n)}{4} + \frac{W}{2}
(g-1)^2 \int_{0}^{n} dy \frac{y(1-y)}{1+(g^2-1)y}
\quad .
\end{eqnarray}
\end{mathletters}%
A simple integration gives
\begin{mathletters}
\begin{eqnarray}
\overline{d} (g) &=& \frac{g^2}{2(1-g^2)^2} \left[ -(1-g^2)n -
\ln \left( 1 -(1-g^2)n\right) \right]
\label{dbarGWF}
\\[6pt]
\langle \hat{T}(g) \rangle/L &=& -\frac{Wn(2-n)}{4} -
W \left( \frac{g-1}{g+1} \right)
\left[ \left( \frac{n}{2}\right)^2 - \overline{d} (g)
\right]
\quad .
\label{TGWF}
\end{eqnarray}%
\end{mathletters}%
As shown in appendix~\ref{appb}
these results completely agree with those obtained from a direct
application of the methods developed in Refs.~\cite{MV,DVFG},
and provides an independent check for the conjectured solution
of the $1/r$-Hubbard model in Ref.~\cite{prl}.

To obtain the variational ground state energy density one has to minimize
$e_0^{\rm G} (g) = \langle \hat{T} (g) \rangle/L + U \overline{d} (g)$
with respect to~$g$. This has to be done numerically
for general~$U/W$. The resulting curve at half-filling is shown in
figure~\ref{fig2}. It is seen that the Gutz\-willer wave function is
very good for values of $U/W \leq 0.5$. Indeed, as shown in
appendix~\ref{appb}, the Gutzwiller wave function gives the exact
ground state energy to order~$U^2/W$ for all fillings.
It is further exact to order~$n^3$ for all values of~$U/W$, see
appendix~\ref{appb}. The logarithmic dependence of~$\overline{d} (g)$
on~$g$ for small~$g$ (i.e., large~$U/W$) causes the variational
energy to behave like~$e_0^{\rm G} \sim W^2/\left[U \ln(U/W)\right]$,
see Ref.~\cite{MV}, in contrast to the exact result, see eq.~(\ref{e0exact}).
As discussed in detail in Ref.~\cite{review}, this wrong behavior
of the variational ground state energy comes from the poor
correlations between double occupancies and holes
in the Gutzwiller wave function for large~$U/W$.

It is clearly seen that $g>0$ for all $U<\infty$ such that the
variational prediction for the metal-to-insulator transition
is $U_c^{\rm G} = \infty$. It should be clear, however, that
at least the physics of this transition is correct: it is due to
charge correlations, magnetic ordering neither occurs nor is it necessary.
Furthermore, the transition only happens at~$n=1$ as it should,
and not for arbitrary electron density as in the case of
Hartree-Fock theory.

\subsection{Baeriswyl wave function}
\label{Baer}
We know that the Gutzwiller-projected Fermi-sea is the exact ground state
of the~$1/r$-tJ model with pair-hopping terms~\cite{prl,prb}, i.e.,
of the Hamiltonian which one gets from eq.~(\ref{hamilt}) by
first order perturbation theory in~$W/U$.
Consequently, the Baeriswyl wave function
is the exact ground state of the $1/r$-Hubbard model~(\ref{hamilt})
to order~$W/U$~\cite{Baeriswyl,review}.

For $g=0$ the equations~(\ref{alphabeta})
can be simplified to $\alpha (x;b) = ( 1-x )/(2\sqrt{b})$, and
$\beta (x;b) = -\sqrt{b}\sqrt{1-x^2}/2$.
After some straightforward manipulations eqs.~(\ref{resultVWF})
finally give
\begin{mathletters}
\begin{eqnarray}
\langle \hat{T}(b) \rangle/L &=&
-\frac{Wn(2-n)}{4} + W \frac{b^2}{2(1-b^2)^2}
\left[ - (1-b^2)n - \ln \left( 1 -(1-b^2) n\right) \right]
\label{TBWF} \\[6pt]
\overline{d} (b) &=& \frac{1-b}{1+b}
\left[ - \frac{\langle \hat{T} \rangle /L}{W} - \frac{n(1-n)}{2} \right]
\label{dbarBWF}
\quad .
\end{eqnarray}%
\end{mathletters}%
Comparing eqs.~(\ref{dbarGWF}) and (\ref{TBWF}), and
eqs.~(\ref{TGWF}) and (\ref{dbarBWF}), we note that the mean kinetic
energy and the mean double occupancy just change their roles when
we go from the Gutzwiller wave function to the Baeriswyl wave function.
However, this only happens in the highly symmetric~$1/r$-Hubbard model.

Form this equivalence it immediately follows
that there is a metal-to-insulator transition for~$n=1$
in the Baeriswyl wave function at~$U_c^{\rm B}=0^+$, i.e.,
this wave function is insulating at half-filling for all~$U>0$.
While for large~$U/W$ we obtain the exact result~$e_0^{B}=
-W^2/(12 U) + {\cal O} \bigl( W^3/U^2 \bigr)$,
the wave function gives only a poor estimate for the ground state
energy density at small~$U/W$. One obtains the Hartree contribution
to first order, but there is, at half-filling, no contribution
to second order in~$(U/W)^2$. Instead, the correction is
proportional to~$(U^2/W) \ln (U/W)$.

At least for the~$1/r$-Hubbard model we can thus conclude that
it is as insufficient as the Gutzwiller wave function in the
opposite regime of its obvious applicability.
This unsatisfactory behavior might in part be due to the special model.
Unfortunately, this wave function cannot be evaluated for other
Hamiltonians away from~$b \lesssim 1$.

\subsection{Gutzwiller-Baeriswyl wave function}
\label{GutzBaer}
For finite~$b$ and~$g$, the Gutzwiller-Baeriswyl wave function represents
a metallic state. We first analyze whether we can have a metal-insulator
transition at a finite value of~$U/W$.

The integrals in~eqs.~(\ref{resultVWF}) can be done analytically.
To properly take care of possible singularities we write
\begin{mathletters}%
\label{TdVWF}
\begin{eqnarray}%
\langle \hat{T} (b,g) \rangle/L &=& -\frac{Wn(2-n)}{4}+\frac{W}{2}
b^2 (g-1)^2 \left( I^{(1)} - I^{(2)} \right) \\[6pt]
2 \overline{d}(b,g) &=&
\left[ 1 + b(g-1)\right]^2 \left( I^{(1)} - I^{(2)} \right)
+ (g-1)^2 (1-b)^2 \left( I^{(3)} - I^{(2)} \right)
+ g^2 I^{(2)}
\end{eqnarray}%
\end{mathletters}%
where
\begin{equation}
I^{(m)} = \int_0^n dy \, \frac{y^m}{A(b,g)y^2+B(b,g)y+1}
\end{equation}
with~$A(b,g)=(g-1)^2(1-b^2) \geq 0$, $B(b,g)=2(g-1)+b^2(g-1)^2 \leq 0$.
The denominator of the integrals~$I^{(m)}$ has zeros
at~$z_{1,2} = (|B| \pm \sqrt{B^2-4A})/(2A)$.
One can easily show that $z_1\geq z_2 \geq 1$. This implies
that there will be a singularity (metal-insulator transition)
only at half-filling (the upper limit of integration is~$n \leq 1$),
if~$z_2=1$. This can only be fulfilled, if~$A(b,g) + B(b,g) +1 =0$,
or~$g=0$, in agreement with our general considerations.

We have to find out whether the variational procedure gives~$g=0$ at
some finite~$U_c^{\rm GB}$.
To this end we have to carefully
analyze the limit~$g\to 0$. We can assume that
$b$ will not be unity because this would correspond
to the case~$U_c^{\rm GB}=\infty$. In the limit~$g\to 0$
we first note that $z_2 \to 1 + {\cal O}(g^2)$, $z_1 \to 1/(1-b^2)$.
Then, the combinations~$I^{(1)} - I^{(2)}$ and~$I^{(3)} - I^{(2)}$
stay finite, and actually are of the order~$g^2\ln g$.
Also, $I^{(2)}(b,g) \sim \ln g$ for small $g$.
If we now expand $e_0^{\rm GB} (b,g)$ around~$g=0$, one finds
\begin{equation}
e_0^{\rm GB} (b,g) = e_0^{\rm B} (b) +
e_1 (b) g + e_2 (b) g^2 \ln g + \ldots \quad .
\label{minimize}
\end{equation}
The first term in this expansion,
$e_1(b) = \left. \partial e_0^{\rm GB} (b,g)/ \partial g\right|_{g=0}$
exists because all singularities
at least of the order~$g^2 \ln g$. After some lengthy calculations
we find
\begin{equation}
e_1(b) = - \frac{W}{(1-b^2)^2}
\left[ 1 -b^2 + 2b^2\ln b\right]
+ U \left[ \frac{1-b^2 -2b\ln b}{(1+b)^3}\right]
\quad .
\end{equation}
Close to the transition we can minimize eq.~(\ref{minimize})
with respect to~$g$ which gives
\begin{equation}
g \ln g = - \frac{e_1(b)}{2 e_2(b)}
\quad .
\label{geq}
\end{equation}
Thus we know~$g(b)$ as a function of~$b$. We insert this into our
expansion~(\ref{minimize}) and obtain
\begin{equation}
e_0^{\rm GB} (b) = e_0^{\rm B}(b) + \frac{g(b)}{2} e_1 (b)
\end{equation}
For given~$U/W$ we have to compare the minima of~$e_0^{\rm GB}(b)$
at~$b_0^{\rm GB}$ and of~$e_0^{\rm B}(b)$ at~$b_0^{\rm B}$.
If we denote the difference in the variational
ground state energy densities by $\Delta e_0= e_0^{\rm GB}(b_0^{\rm GB})
- e_0^{\rm B}(b_0^{\rm B})$, and let~$b_0^{\rm GB} =b_0^{\rm B} +\delta b$
we may write for small~$g$
\begin{eqnarray}
\Delta e_0 &\approx & e_0^{\rm B}(b_0^{\rm GB}) + \frac{g(b_0^{\rm GB})}{2}
e_1(b_0^{\rm GB})
- e_0^{\rm B}(b_0^{\rm B}) \nonumber\\
 & \approx &
 \left. \frac{\partial e_0^{\rm B}(b) }{ \partial b}
 \right|_{b_0^{\rm B}} (\delta b)
 + \frac{1}{2} \left. \frac{\partial^2 e_0^{\rm B}(b)}{\partial b^2}
 \right|_{b_0^{\rm B}}
 (\delta b)^2 + \frac{g(b_0^{\rm GB})}{2}
e_1(b_0^{\rm GB})
\\
& = & \frac{1}{2} \left. \frac{\partial^2 e_0^{\rm B}(b)}{\partial b^2}
 \right|_{b_0^{\rm B}}
 (\delta b)^2 + \frac{g(b_0^{\rm GB})}{2} e_1(b_0^{\rm GB}) \quad .
 \nonumber
\end{eqnarray}
The first term in the expansion in~$\delta b$
vanished by definition of the variational minimum. Accordingly,
the second derivative is positive. The energy~$e_1(b_0^{\rm GB})$
must be negative to allow for~$g>0$ in eq.~(\ref{geq}).

Now we are in the position to decide from the results of the numerical
minimization procedure whether~$\Delta e_0$ is bigger or less than zero.
We essentially find that~$g(b_0^{\rm GB}) = {\cal O}(\delta b)$,
typically $g = 4\cdot 10^{-5}$, $e_1(b_0^{\rm GB}) = - {\cal O}(10^{-2}W)$,
$\delta b = 5 \cdot 10^{-5}$, and
$(\partial^2 e_0^{\rm B})/(\partial b^2) (b_0^{\rm B}) = {\cal O}(W)$
for~$U/W=10$. Hence, the negative term~$g (b_0^{\rm GB})
e_1(b_0^{\rm GB})$ easily overcomes the positive
contribution~$(\delta b)^2 (\partial^2 e_0^{\rm B})/(\partial b^2)
(b_0^{\rm B})$ by
several orders of magnitude. For all~$U/W$ we have~$\Delta e_0 < 0$,
i.e., the Gutzwiller-Baeriswyl wave function for {\em finite}~$g$
always wins over the Baeriswyl wave function ($g=0$).
The metal-to-insulator transition again happens at
$U_c^{\rm GB}=\infty$.
We had to do this detailed analysis because the energies of
the Baeriswyl and the Gutzwiller-Baeriswyl energy become
numerically indistinguishable for values of~$U/W>5$.

In general,
the variational ground state energy density is
obtained from a numerical minimization of
$e_0^{\rm GB} (b,g) = \langle \hat{T} (b,g) \rangle/L + U \overline{d} (b,g)$
with respect to~$b$ and~$g$. Given the fact that the Gutzwiller-Baeriswyl wave
function becomes exact both for small and large~$U/W$ we obtain
an excellent estimate for the ground state energy for all~$U/W$.
We thus plot the {\em relative deviation}
$|(e_0^{\rm GB} -e_0)/e_0|$ from the exact ground state
energy in figure~\ref{fig3}. The maximum error is very small,
about~1.6\% around~$U/W=1$, and strongly decreases for larger and
smaller values of the interaction.

The Gutzwiller-Baeriswyl wave function becomes {\em exact}
both for small and for large~$U/W$, and it gives an excellent estimate
for the exact ground state energy density. However, it still {\em misses}
the metal-to-insulator transition in the~$1/r$-Hubbard model at~$U_c=W$.

\section{Conclusions}
\label{sec4}
In this work we investigated variational approaches
to the exactly solvable~$1/r$-Hubbard Hamiltonian.
We found a two-parameter trial state, the Gutzwiller-Baeriswyl
wave function, which becomes exact in both the weak and strong-coupling
limit, and gives an excellent upper bound for the exact ground state
energy. Yet it {\em fails} to reproduces the metal-to-insulator
transition in the half-filled~$1/r$-Hubbard model at~$U_c=W$.
Instead it predicts~$U_c^{\rm GB}=\infty$, just like the
Gutzwiller wave function, $U_c^{\rm G}=\infty$. On the other hand,
the Baeriswyl wave function predicts~$U_c^{\rm B}=0^+$, also
in contradiction to the exact result.

We thus see that even very elaborate correlated wave functions are
{\em unable} to reproduce the correct ground state physics
of a one-dimensional model. However,
the Gutzwiller-Baeriswyl wave function has to be considered
an ``excellent'' wave function according to the ``energy criterion''.
This implies that the ``energy criterion'' is not very valuable,
if one wants to draw conclusions about the ground state physics
of a given Hamiltonian.
This finding is supported by the fact that the
results of the Gutzwiller Approximation to the Gutzwiller wave function
are in {\em qualitative} agreement with the ground state physics
of the~$1/r$-Hubbard model even though there are considerable
{\em quantitative}  differences. The energy estimate is rather poor,
and not even variationally controlled, but the basic physical
concept that charge-correlations drive the transition
at a finite interaction strength is properly
incorporated in the approximation.

Antiferromagnetic Hartree-Fock theory always overestimates the
role of spin-correlations in the metal-to-insulator
transition while charge-correlations are neglected.
Thus, there is no correlation energy
for small interactions, the transition takes place for all
particle fillings, and a finite sublattice magnetization occurs.
All three features are not present in the exact solution, in part due
to the low dimensionality of our system.
The physics of the metal-to-insulator transition which
is predicted to happen at a finite~$U_c^{\rm HF}=W/2$ is solely a
consequence of spin-correlations, contrary to the physics of
the transition in the model. We know that charge-correlations
drive the Mott-Hubbard metal-to-insulator transition, and
spin-correlations are only residual effects. Thus,
the Hartree-Fock picture is always a qualitatively wrong
description of correlated Fermi systems.

The situation in higher dimensions is more promising
for variational approaches. A successful theory of normal fluid~$^{3}$He
is based on the Gutzwiller wave function~\cite{Vollhardtreview},
evaluated in the limit of high dimensions
where the Gutzwiller Approximation to the Gutzwiller wave function
becomes exact~\cite{MV,MVlarged}.
In low dimensions, however, systems are Luttinger Liquids~\cite{HaldaneLL}
rather than Fermi liquids. Charge-spin separation in such systems
makes it difficult to write down
qualitatively correct wave functions in terms of the original
Fermion operators. Since charge-spin separation is common to many
one-dimensional systems, it is doubtful whether one can write down
successful Jastrow-Feenberg wave functions for correlated
itinerant Fermion systems in low dimensions.

\appendix
\section{Calculation of the ground state energy for correlated wave functions}
\label{appa}
In ref.~\cite{prl} we introduced an effective Hamiltonian description
of the energy spectrum of the~$1/r$-Hubbard Hamiltonian.
We choose a basis in which~$\hat{T}$ is diagonal, and
represent states by putting hard-core bosons for
spin ($\hat{s}_{{\cal K}\sigma}$)
and charge ($\hat{d}_{\cal K},\hat{e}_{\cal K}$)
degrees of freedom onto each site~${\cal K}$, ${\cal K} = \Delta (m+1/2)$
($m=-L/2,\ldots,L/2 -1)$. They obey a hard core constraint,
$\sum_{\sigma} \hat{n}_{{\cal K},\sigma}^{s}
 + \hat{n}_{{\cal K}}^{d} + \hat{n}_{{\cal K}}^{e} = 1$
for each~${\cal K}$. Each site is then occupied by one and
only one of these four objects, $\uparrow$, $\downarrow$, $\bullet$,
$\circ$~\cite{note}.

In this representation the Fermi-sea reads
\[
| \text{Fermi-sea} \rangle \equiv |
[\uparrow \downarrow] \ldots [\uparrow \downarrow]
\biggr|_{{\cal K}=0}
[\bullet \circ] \ldots [\bullet \circ]
\biggr|_{{\cal K}_F}
\circ \ldots \circ
\rangle
\]
where we assumed an even particle number, and ${\cal K}_F= \pi(2n-1)$.
To calculate the energy expectation value
of the Gutzwiller-Baeriswyl wave function
we only need to know how~$\hat{T}$ and~$\hat{D}$ act on the
Fermi sea. From Ref.~\cite{prl} we know that the kinetic energy
operator can be split into two parts,
\begin{mathletters}%
\begin{eqnarray}%
\hat{T}^{\rm eff}_{0} & = & \mathop{{\sum}'}_{-\pi < {\cal K} <
\pi\phantom{\prime}}
\left\{ \frac{t{\cal K}}{2} \left[ \left( \hat{s}_{{\cal K},\uparrow}^{+}
\hat{s}_{{\cal K},\uparrow}^{\phantom{+}} + \hat{s}_{{\cal K},\downarrow}^{+}
\hat{s}_{{\cal K},\downarrow}^{\phantom{+}} \right)
-
\left( \hat{d}_{{\cal K}}^{+} \hat{d}_{{\cal K}}^{\phantom{+}}
+ \hat{e}_{{\cal K}}^{+}
\hat{e}_{{\cal K}}^{\phantom{+}} \right) \right]
\right\}
\label{firstT} \\[6pt]
\hat{T}^{\rm eff}_{1} & = &
\sum_{-\pi < {\cal K} < \pi-\Delta} \lambda_{{\cal K}} \pi t
\left[
\hat{s}_{{\cal K},\uparrow}^{+} \hat{s}_{{\cal K},\uparrow}^{\phantom{+}}
\hat{s}_{{\cal K}+\Delta,\downarrow}^{+}
\hat{s}_{{\cal K}+\Delta,\downarrow}^{\phantom{+}}
-
\hat{d}_{{\cal K}}^{+} \hat{d}_{{\cal K}}^{\phantom{+}}
\hat{e}_{{\cal K}+\Delta}^{+}
\hat{e}_{{\cal K}+\Delta}^{\phantom{+}}
\right]
\label{secondT}
\end{eqnarray}%
\end{mathletters}%
where $\lambda_{\cal K}= {\rm sgn}({\cal K})$, and
the prime on the sum in~$\hat{T}^{\rm eff}_{0}$ indicates
that it must not be applied to the states represented by
$[\uparrow \downarrow]$ and $[\bullet \circ]$~\cite{prl} which
are treated separately by~$\hat{T}^{\rm eff}_{1}$.
The total effective representation of the kinetic energy operator is
$\hat{T}^{\rm eff} = \hat{T}^{\rm eff}_0 + \hat{T}^{\rm eff}_1$.

The double occupancy operator is expressed by the
two terms
\begin{mathletters}%
\begin{eqnarray}%
\hat{D}^{\rm eff}_{0} & = & \mathop{{\sum}'}_{-\pi < {\cal K} <
\pi\phantom{\prime}}
 \hat{d}_{{\cal K}}^{+} \hat{d}_{{\cal K}}^{\phantom{+}}
\label{firstD}
\\[6pt]
\hat{D}^{\rm eff}_{1} & = &
\sum_{-\pi < {\cal K} < \pi-\Delta}
\Biggl\{
\left[ \frac{1}{2} +\lambda_{{\cal K}} \frac{1}{4\pi}
\left(2{\cal K} + \Delta \right)  \right]
\hat{d}_{{\cal K}}^{+} \hat{d}_{{\cal K}}^{\phantom{+}}
\hat{e}_{{\cal K}+\Delta}^{+}
\hat{e}_{{\cal K}+\Delta}^{\phantom{+}}\nonumber \\
& & \phantom{\sum_{-\pi < {\cal K} < \pi-\Delta} \biggl\{ }
+ \left[ \frac{1}{2} - \lambda_{{\cal K}}  \frac{1}{4\pi}
\left(2{\cal K} + \Delta \right)  \right]
\hat{s}_{{\cal K},\uparrow}^{+} \hat{s}_{{\cal K},\uparrow}^{\phantom{+}}
\hat{s}_{{\cal K}+\Delta,\downarrow}^{+}
\hat{s}_{{\cal K}+\Delta,\downarrow}^{\phantom{+}}
\label{secondD} \\
& &  \phantom{\sum_{-\pi < {\cal K} < \pi-\Delta} \biggl\{ }
+ \frac{1}{4\pi} \sqrt{(2\pi)^{2} - (2{\cal K}+\Delta)^{2}}
\left[ \hat{d}_{{\cal K}}^{+} \hat{e}_{{\cal K}+\Delta}^{+}
\hat{s}_{{\cal K},\uparrow}^{\phantom{+}}
\hat{s}_{{\cal K}+\Delta,\downarrow}^{\phantom{+}} + {\rm H.c.} \right]
\Biggr\} \; .
\nonumber
\end{eqnarray}%
\end{mathletters}%
The total effective double occupancy operator is
$\hat{D}^{\rm eff} = \hat{D}^{\rm eff}_0 + \hat{D}^{\rm eff}_1$.

Fortunately, we only need to apply
$\hat{T}^{\rm eff}$ and $\hat{D}^{\rm eff}$ to the Fermi-sea when we
calculate the ground state energy of the Gutzwiller-Baeriswyl
wave function. The states for~${\cal K} > {\cal K}_F$ are already diagonal
in~$\hat{T}$ and~$\hat{D}$, and just give an additive constant.
We are left with the action of~$\hat{T}_1^{\rm eff}$
and~$\hat{D}_1^{\rm eff}$ which only act on
neighboring {\em pairs}~$[\uparrow \downarrow]$, and~$[\bullet \circ]$
on~${\cal K}$, ${\cal K}+\Delta$.
We indicate this by a prime on the~${\cal K}$-sums.
For our purposes we may thus use the following matrix
representation for~$\hat{T}^{\rm eff}$
\begin{mathletters}%
\begin{eqnarray}
\hat{T}^{\rm eff} & = & \mathop{{\sum}'}_{-\pi < {\cal K} <
{\cal K}_F-\Delta \phantom{\prime}}
\tensor{T}_{{\cal K}, {\cal K}+\Delta}
+ \sum_{ {\cal K}_F < {\cal K} < \pi}  \left( - \frac{t {\cal K} }{2} \right)
\\[6pt]
\tensor{T}_{{\cal K}, {\cal K}+\Delta} & = & t \pi \left(
\begin{array}{cc}
\lambda_{{\cal K}} & 0 \\
0 & - \lambda_{{\cal K}}
\end{array}
\right)
\end{eqnarray}%
\end{mathletters}%
and for~$\hat{D}^{\rm eff}$
\begin{mathletters}%
\begin{eqnarray}
\hat{D}^{\rm eff}  & = & \mathop{{\sum}'}_{-\pi < {\cal K} <
{\cal K}_F-\Delta \phantom{\prime} } \tensor{D}_{{\cal K}, {\cal K}+\Delta}
\\[6pt]
\tensor{D}_{{\cal K}, {\cal K}+\Delta} & = &
\frac{1}{2}
 \left(
 \begin{array}{cc}
1- \lambda_{{\cal K}} a_{\cal K} & \sqrt{1-a_{\cal K}^2} \\
\sqrt{1-a_{\cal K}^2} & 1+ \lambda_{{\cal K}} a_{\cal K}
\end{array}
\right)
\end{eqnarray}%
\end{mathletters}%
where $a_{\cal K}= (2{\cal K}+\Delta)/(2\pi)$.
The matrix~$\tensor{D}_{{\cal K}, {\cal K}+\Delta}$ has the eigenvalues
zero and one.

We see that the problem of calculating expectation values~$\langle \hat{T}
\rangle$ and~$\langle \hat{D} \rangle$ factorizes into independent
$2\times 2$ matrix problems.
We obtain
\begin{mathletters}%
\label{TDmatrices}
\begin{eqnarray}%
\langle \hat{T} \rangle &=&
\sum_{{\cal K}_F < {\cal K} < \pi} \left( - \frac{t {\cal K} }{2} \right)
+ \mathop{{\sum}'}_{-\pi < {\cal K} < {\cal K}_F-\Delta \phantom{\prime}}
\frac{%
\loarrow{\Lambda}_{{\cal K}, {\cal K}+\Delta}
\tensor{T}_{{\cal K}, {\cal K}+\Delta}
\roarrow{\Lambda}_{{\cal K}, {\cal K}+\Delta}
}%
{%
\loarrow{\Lambda}_{{\cal K}, {\cal K}+\Delta}
\roarrow{\Lambda}_{{\cal K}, {\cal K}+\Delta}
}
\\[6pt]
\langle \hat{D} \rangle &=&
\mathop{{\sum}'}_{-\pi < {\cal K} < {\cal K}_F-\Delta \phantom{\prime}}
\frac{%
\loarrow{\Lambda}_{{\cal K}, {\cal K}+\Delta}
\tensor{D}_{{\cal K}, {\cal K}+\Delta}
\roarrow{\Lambda}_{{\cal K}, {\cal K}+\Delta}
}%
{%
\loarrow{\Lambda}_{{\cal K}, {\cal K}+\Delta}
\roarrow{\Lambda}_{{\cal K}, {\cal K}+\Delta}
}
\quad .
\end{eqnarray}%
\end{mathletters}%
$\loarrow{\Lambda}_{{\cal K}, {\cal K}+\Delta}$ is the transpose
of the vector
\begin{equation}
\roarrow{\Lambda}_{{\cal K}, {\cal K}+\Delta}
=
\tensor{M}_{{\cal K}, {\cal K}+\Delta}
\left\{
\begin{array}{ccl}
\left( \begin{array}{c} 1 \\[-5pt] 0 \end{array} \right)
& \text{for} & -\pi < {\cal K} < 0 \\[6pt]
\left( \begin{array}{c} 0 \\[-5pt] 1 \end{array} \right)
& \text{for} & 0 < {\cal K} < {\cal K}_F
\end{array}
\right.
\end{equation}
with the ``correlation matrix'' ($\tensor{1}$ is the $2\times 2$ unit matrix)
\begin{eqnarray}
\tensor{M}_{{\cal K}, {\cal K}+\Delta} &=&
b^{\tensor{T}_{{\cal K}, {\cal K}+\Delta}/W}
g^{\tensor{D}_{{\cal K}, {\cal K}+\Delta}} \nonumber \\
 & = &
 \left(
\begin{array}{cc}
b^{\lambda_{{\cal K}}/2} & 0 \\
0 & b^{-\lambda_{{\cal K}}/2}
\end{array}
\right)
\left( \tensor{1} + (g-1) \tensor{D}_{{\cal K}, {\cal K}+\Delta}\right)
\quad .
\end{eqnarray}
We thus get~$\loarrow{\Lambda}_{{\cal K}, {\cal K}+\Delta}
= ( \alpha (a_{{\cal K}};b,g) \; \beta (a_{{\cal K}};b,g)$
for~$-\pi < {\cal K} < 0$,
and $\loarrow{\Lambda}_{{\cal K}, {\cal K}+\Delta}
= ( \beta (a_{{\cal K}};b,g) \; \alpha (a_{{\cal K}};b,g)$
for~$0 < {\cal K} < {\cal K}_F$ where
$\alpha (x;b,g)$ and $\beta (x;b,g)$ were defined in
eqs.~(\ref{alphabeta}).

Going from ${\cal K} < 0$ to ${\cal K}>0$ we see that the two entries
in $\loarrow{\Lambda}_{{\cal K}, {\cal K}+\Delta}$ are exchanged.
The same happens to row and columns in the Hamilton matrix
$\tensor{H}_{{\cal K}, {\cal K}+\Delta} = \tensor{T} + U \tensor{D}$
such that we may drop the
distinction between the two cases.
It now is a straightforward task to evaluate
eqs.~(\ref{TDmatrices}) in terms of
$\alpha (x;b,g)$ and $\beta (x;b,g)$. In the
thermodynamical limit one easily arrives at eq.~(\ref{resultVWF}),
if one takes into account that the primed sums give an extra
factor of a half in front of the integral because only every second
${\cal K}$-value has to be included.

\section{Alternative Calculation of the ground state energy for the Gutzwiller
wave function}
\label{appb}
We want to directly calculate the ground state energy of the Gutzwiller
wave function by using the methods of Vollhardt et al.~\cite{MV,DVFG}.
To this end we first shift the Fermi-sea of filled states between
$k=-\pi$ and $k_F^e=\pi(n-1)$ by $(\pi-k_F^e)/2$ to generate a symmetric
Fermi body~$\langle \widetilde{n}_{k,\sigma} \rangle$.
This transformation does not change the double occupancy operator.
We may then directly apply all formulae of Refs.~\cite{MV,DVFG}.
For example, the formula for the average double occupancy~$\overline{d}(g)$
agrees with eq.~(\ref{dbarGWF}). To obtain the average kinetic energy
we have to calculate~$\sum_{k,\sigma} \epsilon\left[ k -(\pi-k_F^e)/2 \right]
\langle \widetilde{n}_{k,\sigma} \rangle$ where
$\langle \widetilde{n}_{k,\sigma} \rangle$ can recursively
be obtained~\cite{MV}. In our case we do not need such
detailed information to calculate the average kinetic energy.
We take the lattice periodicity of our dispersion relation into account,
and write
\begin{equation}
\langle \hat{T}(g) \rangle = - \frac{t}{2} (\pi -k_F^e) \sum_{-\pi < k <
\pi,\sigma}
\langle \widetilde{n}_{k,\sigma} \rangle +
2\pi t \sum_{-\pi < k < -\pi+(\pi-k_F^e)/2}
\langle \widetilde{n}_{k,\sigma} \rangle
\label{singleeq}
\end{equation}
where we dropped the term
$\sum_{-\pi< k < \pi,\sigma} tk \, \langle \widetilde{n}_{k,\sigma} \rangle =0$
which vanishes because the dispersion is antisymmetric
but the shifted momentum distribution~$\langle \widetilde{n}_{k,\sigma}\rangle$
is symmetric around~$k=0$.
The first~$k$-sum in~(\ref{singleeq}) gives the total number of electrons
while the second gives half the number of electrons outside the
Fermi body,~$n_{\rm out}/2$, for which a closed expression
can be given~\cite{MV}
\begin{equation}
n_{\rm out} = 2 \frac{1-g}{1+g} \left[ \left( \frac{n}{2}\right)^2
-\overline{d}(g) \right] \quad .
\end{equation}
The average kinetic energy then indeed
reduces to eq.~(\ref{TGWF}).

We will now show that the Gutzwiller wave function gives the
exact ground state energy to order~$(U/W)^2$.
To this end we recall~\cite{prl,prb} that the energy spectrum of the kinetic
energy operator for fixed total momentum
consists of equidistant levels, separated by~$W$. The potential energy
operator can be split into three parts: $\hat{D}_0$ which commutes
with~$\hat{T}$, and $\hat{D}_{\pm}$ which scatter eigenstates of~$\hat{T}$
such that the final state has a by~$W$ higher (lower) kinetic energy.
To first order in~$U/W$ we can write the (non yet normalized)
exact ground state as
$|\psi_0 \rangle = \left[ 1 + \alpha \hat{D} \right] |\hbox{Fermi-sea}\rangle$
with $\alpha = {\cal O} (U/W)$ because the Fermi-sea is
the unique ground state (the application of $\hat{D}_0$ only gives
the Fermi-sea again). This expression is nothing but the Gutzwiller
wave function to order~$(g-1)$ such that the two wave functions agree
to first order in perturbation theory (the minimization procedure gives
$1-g=U/W$). Consequently, the Gutzwiller wave function gives the
exact ground state energy to order~$(U/W)^2$ for the $1/r$-Hubbard model.

A similar line of arguments can be used to show that the
Gutzwiller wave function also gives the exact ground state energy
to order~$n^3$ for all~$U/W$. To this end, we recall that
the exact ground state wave function for small densities
is given by the Fermi sea and
{\em one} particle-hole pair which is created by the action of~$\hat{D}_+$.
For low densities only ladder-diagrams contribute which
means that no further particle-hole pairs are created but only
the existing one is scattered. This corresponds to the
repeated action of~$\hat{D}_0$ in the subspace of the {\em same} kinetic
energy.
This means that all eigenstates of the kinetic energy operator
which correspond to one particle-hole pair retain the same
phase factor.
To first order in~$n$ we may replace all operators~$\hat{D}_{0}$,
$\hat{D}_{\pm}$ by~$\hat{D}$. It is thus seen that the exact wave function
to first order in~$n$ has the same states
with one particle-hole pair as the Gutzwiller wave function.
In particular, these states have the same relative phases.
Consequently, the Gutzwiller wave function
is exact to the first non-trivial order in the density, and thus
gives the exact ground state energy to order~$n^3$ for all interaction
strength~$U/W$.
The corresponding optimal value of~$g$ is~$g=1/(1+U/W)$.

We see that lowest-order perturbation theory for the $1/r$-Hubbard model
is easily done because the Gutzwiller wave function becomes
exact in this limit.

\begin{figure}
\caption{Hartree-Fock ($e_0^{\rm HF}$) and exact ($e_0$)
ground state energy density
at half-filling as function of $U/W$.
For comparison we also include the result of the Gutzwiller Approximation
to the Gutzwiller wave function, $e_0^{\rm GA}$.}
\label{fig1}
\end{figure}

\begin{figure}
\caption{Exact ($e_0$) and Gutzwiller/Baeriswyl ($e_0^{\rm G}$,
$e_0^{\rm B}$) variational ground state energies
as function of~$U/W$ for the half-filled case; on this scale
there is no distinction possible between the exact ground
state energy density and the variational result from
the two-parameter Gutzwiller-Baeriswyl wave function, $e_0^{\rm GB}$.}
\label{fig2}
\end{figure}

\begin{figure}
\caption{Relative error~$|(e_0^{\rm GB} -e_0)/e_0|$
of the Gutzwiller-Baeriswyl variational
energy, $e_0^{\rm GB}$, compared to the exact ground state
energy density, $e_0$, at half-filling.}
\label{fig3}
\end{figure}

\begin{references}
\bibitem{note} In ref.~\cite{prl,prb} we used this pictorial representation
for eigenstates of the effective Hamiltonian while we use it here
for the representation of eigenstates of the kinetic energy operator.
\end{references}
\end{document}